\documentclass[letterpaper,10pt,oneside]{article}
\usepackage[dvips]{color}
\usepackage{graphicx}
\usepackage[scanall]{psfrag}
\usepackage{natbib}
\renewcommand{\baselinestretch}{1}

\begin{document}

\begin{titlepage}

\large

\LARGE
\centerline{Stability and the Evolvability}
\centerline{of Function in a Model Protein}
\bigskip

\large
\centerline{Running Title: Evolving Stability and Function} 

\bigskip

\centerline{Jesse D. Bloom~\footnotemark[1]~$^{\mbox{\small \#}}$~\footnotemark[4]~, Claus O. Wilke~$^{\mbox{\small \#}}$, 
Frances H. Arnold\footnotemark[5]~, and Christoph Adami~$^{\mbox{\small \#}}$~\footnotemark[6]}
\bigskip

\centerline{\footnotemark[1] Department of Chemistry, $^{\mbox{\small \#}}$ Digital Life Laboratory, 136-93}

\centerline{and \footnotemark[5] Division of Chemistry and Chemical Engineering,}

\centerline{California Institute of Technology, Pasadena, CA  91125}

\centerline{and \footnotemark[6] Jet Propulsion Laboratory 126-347, California Institute}

\centerline{of Technology, Pasadena, CA 91109}

\bigskip

\centerline{\footnotemark[4] Corresponding author: California Institute of Technology}

\centerline{Mail Code 210-41}

\centerline{Pasadena, CA  91125}

\centerline{Phone: 626-354-2565, Fax: 626-568-8743, E-mail: bloom@caltech.edu}
\bigskip

\bigskip

\centerline{Keywords: sequence space, lattice model, protein evolution,}

\centerline{thermostability, thermodynamic stability, directed evolution}

\end{titlepage}

\begin{abstract}
Functional proteins must fold with some minimal stability to a structure
that can perform a biochemical task.
Here we use a simple model to 
investigate the relationship between the stability requirement and 
the capacity of a protein to evolve the function of binding to a ligand.  
Although our model contains no built-in tradeoff between stability
and function, proteins evolved function 
more efficiently when the stability requirement
was relaxed.  Proteins with both 
high stability and high function evolved more efficiently
when the stability requirement was gradually increased 
than when there was constant selection for high stability.
These results
show that in our model,
the evolution of function is enhanced by allowing proteins
to explore sequences corresponding to marginally stable structures,
and that it is easier to improve stability while maintaining high function
than to improve function while maintaining high stability.  Our model also
demonstrates that even in the absence of a fundamental biophysical tradeoff
between stability and function, the speed with which function can evolve
is limited by the stability requirement imposed on the protein.
\end{abstract}

\section*{}

For nearly all proteins found in nature, there is a
unique mapping from the linear protein sequence to a 
thermodynamically stable three-dimensional
native structure, with the mapping determined by the laws of physics~\citep{Anfinsen}. 
However, this unique mapping from sequence to conformation is not 
a general property of polypeptide sequences, since most randomly generated sequences
do not have stable folded structures~\citep{Keefe,Davidson}.  In other words, natural 
protein sequences exist in the space of 
foldable sequences, which is but a small subset of the
space of all possible sequences.  Therefore, evolution must have acted heavily
on natural proteins in order to select those with stable native structures.  

Although natural proteins possess stable native structures,
the evolutionary fitness of a protein depends not on the stability of the
native structure 
\textit{per se}, but rather on the stability of this structure being appropriate to
allow the protein to perform a function
such as catalyzing a chemical reaction or binding to a ligand.  Stability is
therefore under selection only insofar as it is necessary for biochemical function, and most
natural proteins are only marginally stable at their physiologically
relevant temperatures~\citep{Fershtbook}. 

In protein mutagenesis studies, stability and function can appear to be competing
properties, with
mutations that increase stability
often reducing function~\citep{Shoichet,Schreiber}, and mutations that improve
or alter function often decreasing stability~\citep{Wang}.
However, several lines of evidence demonstrate that high stability
and high functionality are not inherently incompatible.
In Nature, there is a strong correlation between the temperature of
an organism's environment and the stability of its proteins, indicating
that natural evolution is able to create functional and highly stable
proteins if there is sufficiently strong selection 
pressure~\citep{Somero1995,hyperthermophiles}. 

In the laboratory, protein engineers have also demonstrated 
that natural proteins are not maximally stable 
by using directed evolution
to find mutations that make proteins more stable without
sacrificing enzymatic 
function~\citep{Giver,Arnold,Serrano,Arnoldreview}.
These results show that high functionality and high stability can coexist,
suggesting that
the marginal stabilities of natural proteins are due
primarily to the simple fact that highly stable sequences are 
rare~\citep{Taverna2002}, and therefore that 
most mutations to an evolved protein will decrease
its stability.
For this reason, proteins will tend to be no more stable than
is required by their environment, since any extra stability that confers
no further selective advantage will be eliminated by mutations. 

Comprehensive experimental examinations of protein evolution are
limited by the vast number of possible sequences and the 
difficulties in rapidly assaying protein properties.
However, simple protein models originally developed to study
protein folding~\citep{Dillreview,Levitt1994,Shakhnovich1993,Wolynesreview}
provide a useful tool for studying
protein evolutionary dynamics~\citep{latticereview}. 
While these models are gross oversimplifications of real proteins,
their tractability allows for a far more extensive exploration
of sequence space than can be done experimentally. 
Previous studies using model proteins have focused on the evolution of
stable structures~\citep{Xia,Cui,Bastolla,Taverna,Tiana,Bornberg-Bauer}
or fast-folding~\citep{Gutin1995,Mirny1998} proteins, while 
with few exceptions~\citep{latticefunction,Hirst1999} the interplay
betweeen the evolution of stability and function has gone unexamined. 
Here we use a model protein to investigate how selection for stability
affects the evolution of function.
In our model, we describe the function of a protein as its ability to bind to
a rigid ligand molecule.  The fitness of a protein
depends on its ability to perform its function of binding to a ligand,
which in turn depends on its ability to fold to a native structure
with some minimal stability. 
We can increase the minimal stability requirement
by increasing the temperature parameter, allowing us to explore the 
relationship between stability and the evolvability of function.

\section*{Methods}

\subsection*{The Protein Model}

We use a highly simplified model of a protein consisting of a chain of $N = 18$ monomers
on a two-dimensional lattice that we allow to occupy any compact or noncompact
conformation.

The monomers can be of $20$ types, corresponding to the $20$ amino acids. 
Each monomer on the lattice has four nearest neighbor sites, of which as
many as two can be occupied by non-bonded neighboring residues (three in the case of terminal
residues).  The energy $E\left(\mathcal{C}\right)$ of a protein 
conformation $\mathcal{C}$ is the sum of the 
nearest-neighbor interactions of non-bonded residues,  
$$E\left(\mathcal{C}\right) = \sum\limits_{i=1}^N \sum\limits_{j=1}^{i-2} C_{ij}\left(\mathcal{C}\right)
\times \epsilon\left(\mathcal{A}_{i},\mathcal{A}_{j}\right),$$
where $C_{ij}\left(\mathcal{C}\right)$ equals one if residues $i$ and $j$ are nearest neighbors
in conformation $\mathcal{C}$ and zero otherwise, and 
$\epsilon\left(\mathcal{A}_{i},\mathcal{A}_{j}\right)$ is the interaction 
energy between residue types
$\mathcal{A}_{i}$ and $\mathcal{A}_{j}$.  The interaction energies 
$\epsilon\left(\mathcal{A}_{i},\mathcal{A}_{j}\right)$ are based on a 
widely used statistical
analysis of real proteins by Miyazawa and Jernigan 
\citep{Miyazawa}(Table V).
All energies are given in reduced units such that one energy unit equals $k_{B} T$ at room
temperature ($298$ K).  Temperatures are given in units such that 
$T = 1.0$ at room temperature. 

\subsection*{Folding the Proteins}

The native structure and stability of the protein can be determined by finding the 
lowest energy conformation, $\mathcal{C}_{\rm{low}}$, and the partition function.
Computation of the partition function requires defining a temperature parameter $T$.
This temperature parameter represents the thermodynamic temperature, however since
the model protein interaction energies are independent of temperature, the temperature
parameter does not capture behaviors of real proteins that are caused by the temperature
dependence of the interaction energies (for example, cold denaturation).
In order to avoid confusion, we refer to $T$ as the temperature parameter rather than
as the temperature.

The partition function at a temperature parameter of $T$ is:
$$Q\left(T\right) = \sum\limits_{\left\{\mathcal{C}_{i}\right\}} \exp\left[-E\left(\mathcal{C}_{i}\right) / T\right],$$
where the sum is taken over all conformations $\left\{\mathcal{C}_{i}\right\}$.  
The free energy of folding $\Delta G_{f}\left(T\right)$ to $\mathcal{C}_{\rm{low}}$ is 
then the difference between $E\left(\mathcal{C}_{\rm{low}}\right)$ 
and the free energy of the ensemble of all other conformations,
$$\Delta G_{f}\left(T\right) = E\left(\mathcal{C}_{\rm{low}}\right) + T \ln \left\{Q\left(T\right) - \exp\left[-E\left(\mathcal{C}_{\rm{low}}\right) / T\right]\right\}.$$
The fraction of proteins $f\left(T\right)$ 
that are expected to be folded to $\mathcal{C}_{\rm{low}}$ 
at equilibrium is given by
$$f\left(T\right) = \frac{1}{1 + \exp\left[\Delta G_{f}\left(T\right) / T\right]}.$$

Exact calculation of $Q\left(T\right)$ requires enumeration of 
all $5.81 \times 10^{6}$ unique conformations 
corresponding to all of the self--avoiding
walks that are not related by symmetry~\citep{latticestatistics}.
Many of these walks have very few contacts, and so 
make only a small contribution
to the partition function.  
We only explicitly considered the $7.95 \times 10^{5}$ conformations with
more than four contacts.  The remaining $5.01 \times 10^{6}$ conformations
were treated by a crude mean--field model, estimating the partition
sum contribution of all conformations with $n$ contacts ($0 \le n \le 4$)
as 
$$Q_{n}\left(T\right) = \exp\left(\frac{- n \langle \epsilon \rangle}{T}\right) \times
\exp\left(\frac{- n \sigma^{2}_{\epsilon}}{2 T}\right) \times \mathcal{N}\left(n\right)$$ 
where $\langle \epsilon \rangle$ is the average residue--residue contact energy
for the given protein sequence assuming any residue is equally likely to
be in contact with any other non--adjacent residue, $\sigma^{2}_{\epsilon}$
is the variance in the residue--residue contact energy, and $\mathcal{N}\left(n\right)$
is the number of conformations with $n$ contacts. 
This approximation 
introduces only a very small error ---
a test of $10^{3}$ random sequences at 
$T = 1.0$ showed that the root--mean--square error and maximum
differences between the approximate and exact values
of $\Delta G_{f}\left(T\right)$ were
$1.6 \times 10^{-4}$ and $2.8 \times 10^{-3}$ respectively.
This error had no effect on the evolutionary trajectories,
since running a sample trajectory with and without the
approximation led to identical results.
Folding
a protein took roughly $0.3$ seconds on a $2$ GHz processor.

\subsection*{Modeling the Protein Function}

We introduce the concept of function by considering the binding of a 
ligand to the protein, an idea which to our knowledge was first introduced 
by \citep{Miller1997}. 
We define the function of a protein as its ability to bind to a 
rigid ligand when the protein is in
its lowest energy conformation $\mathcal{C}_{\rm{low}}$.  
The binding energy 
$\rm{BE}\left(\mathcal{C}_{\rm{low}}\right)$ is the interaction energy
between the folded protein and the rigid ligand in the lowest energy binding position,
found by searching all translational
and rotational positions of the ligand relative to the protein.
Figure \ref{fig:BindingExample}
shows the ligand used in the simulations bound to a protein in its lowest energy conformation.

\subsection*{The Fitness Function}

The fitness $\mathcal{F}\left(T\right)$ of a protein at a temperature parameter of
$T$ is defined as the 
negative product of the fraction of the proteins
that are folded $f\left(T\right)$ times the binding 
energy $\rm{BE}\left(\mathcal{C}_{\rm{low}}\right)$ of the folded protein to the ligand, so that

\begin{eqnarray*}
\mathcal{F}\left(T\right) & = & - f\left(T\right) \times \rm{BE}\left(\mathcal{C}_{\rm{low}}\right) \\
& = & - \frac{1}{1 + \exp\left[\Delta G_{f}\left(T\right) / T\right]} \times \rm{BE}\left(\mathcal{C}_{\rm{low}}\right).
\end{eqnarray*}

Since $\mathcal{F}\left(T\right)$ has a sigmoidal dependence on $\Delta G_{f}\left(T\right)$, the fitness
of an unstructured protein is essentially zero, and a protein gains fitness
as it achieves some minimal stability determined by the temperature parameter.  
Once a protein has achieved the minimal stability,
it can only substantially improve its fitness by improving its ligand binding function.    
The stringency of the stability requirement depends on the temperature parameter
$T$ at which the fitness is computed, with higher temperature parameters favoring greater stabilities. 

Strictly speaking, the fraction of proteins bound to the ligand also displays a sigmoidal
dependence on the product of the ligand binding energy and the fraction of folded proteins.  However,
since we are only interested in differences in fitnesses rather than their magnitudes, any 
function that monotonically increases with this product will give the same results, and so
we choose the simpler functional form defined above.

\subsection*{Evolving the Proteins}

Each evolutionary replicate began with a population of $99$ 
random sequences.
At each generation, the $33$ most fit sequences were selected, and each
was used to generate two identical offspring.  
Random point mutations were 
made in all $99$ resulting proteins with a per-site mutation rate of $3.3 \times 10^{-2}$, which
corresponds to a per-protein mutation rate of $0.6$.  The mutated proteins
were then re-folded, and their fitnesses were calculated. 

For the evolutionary trajectories that began with evolved proteins, we
first evolved random populations for $250$ generations
to bind to each
of the ligands shown in Figure \ref{fig:ligands}.  The best binding sequences
for ligands one, two, and three were 
FFKFKKFKIFMLKWMKMF, FMGFMIIFFLKFKKFGWF, and
MFHVFCHFEWPKPMKCFM respectively.  These sequences were then used for the 
initial identical populations of $99$ proteins
for the evolutionary runs, which were otherwise
carried out as before. 

\section*{Results}

\subsection*{Rapid Evolution of Fitness at Different Stabilities}

We carried out $50$ evolutionary replicates each beginning with 
a different initial random population at 
five different reduced temperature parameters, $T = 0.8$, $0.9$, $1.0$, $1.1$, and $1.2$.
All replicates exhibited similar evolutionary trajectories, with a rapid gain of
fitness in the first few hundred generations followed by only small subsequent
increases in fitness.  Figure \ref{fig:typical} shows two typical replicates 
at a temperature parameter of $1.0$.  
Both the stability and the binding energy improved over time,
with improvements in binding energy usually associated with temporary small decreases 
in stability.

Figure \ref{fig:typical} also shows the lowest energy conformations at four
different points in the evolutionary trajectory for the
two proteins.  The overall structures of the proteins 
were highly conserved,
and the stabilities and binding energies adjusted primarily by 
changes in sequence that preserved these basic structures.  This behavior
is consistent
with the evolution of real proteins, which is believed to involve 
changes in sequence that conserve the
structural scaffold of the protein.
The ligand binding arrangement shown in Figure \ref{fig:BindingExample} 
is typical of that of
an evolved protein.  The most fit proteins usually evolved to fold to compact
conformations with a binding region that fits into the cavity of the ligand. 

\subsection*{Proteins Evolve Ligand Binding Function More Efficiently at Lower Stability}

In order to determine how selection for stability affected the
evolution of ligand binding function, we examined the binding energies achieved after
$500$ generations of evolution at all $5$ temperature parameters.
Figure \ref{fig:temphist} shows the distribution
of binding energies for all cases.
At least a few replicates evolved strong binding proteins at all of the 
temperature parameters.  However, the frequency of evolution of strong binders
was much higher at lower temperature parameters, whereas at higher temperature parameters 
many of the evolutionary trajectories became stuck at weak binding proteins.
The binding energy distributions are statistically different for
temperature parameters that varied by more than $0.2$ with confidences
of greater than $0.95$
(Kolmogorov-Smirnov test, $D$ and $P$ values
for comparison of $T = 0.8$ and $T = 1.0$, 
$T = 0.8$ and $1.1$, $T = 0.8$ and $1.2$, $T = 0.9$ and $T = 1.1$,
$T = 0.9$ and $1.2$, and $T = 1.0$ and $T = 1.2$
are $0.64$ and $7.8 \times 10^{-10}$, $0.58$ and $3.7 \times 10^{-8}$, 
$0.68$ and $4.8 \times 10^{-11}$, $0.34$ and $4.4 \times 10^{-2}$,
$0.60$ and $1.08 \times 10^{-8}$, and $0.52$ and $1.2 \times 10^{-6}$ 
respectively \citep{KStest}).  

Table \ref{tab:stats} shows the mean binding energies, stabilities, and fitnesses evolved by
the proteins.  At higher temperature parameters, the proteins evolved higher stability
but weaker binding.  These results indicate that strong selection for
stability inhibits the evolution of strong binding.
The evolution of a few strong binders at high temperature parameters shows that 
it is fundamentally possible to
evolve good binding under strong selection for stability, however the results
clearly indicate that the statistical likelihood of evolving
strong binding is decreased by increasing the selection for stability.

\subsection*{Low Stability Evolution as a Route to High Stability Fitness}

Since sequences tend to evolve stronger binding at lower temperature parameters,
we speculated that it might be possible to evolve high stability and
strong binding proteins more efficiently by performing the initial generations
at low temperature parameters.  This approach is analogous to simulated annealing, 
except that in
this case the temperature parameter is being increased, since performing
the initial generations at low a temperature parameter helps the proteins escape 
weak binding traps.
We tested this idea by performing
$50$ replicates in which the temperature parameter was set at $0.8$ for the first $200$ steps,
then increased in a linear gradient from $0.8$ to $1.2$ for $200$ steps, and then kept
at $1.2$ for the final $100$ steps.  We compared the average final fitness of the proteins
after this $500$ generation gradient to the final fitness after
$500$ generations at a constantly high temperature parameter of $1.2$ to determine which
approach better allows the proteins to optimize the combination of stability and
ligand binding function.  Figure \ref{fig:grad_fit} shows that the proteins tended to
evolve higher fitness with the gradient than with the constantly
high temperature parameter.  The two distributions are statistically different with 
high confidence
(Kolmogorov-Smirnov test, $D = 0.40$, $P = 4.2 \times 10^{-4}$ \citep{KStest}). 

The gradient approach is more efficient at evolving high fitness because
it prevents the proteins from becoming trapped in regions of high stability but weak binding
by allowing them to first evolve strong binding and then improve their stability.
This can be seen in the two typical trajectories shown in Figure \ref{fig:courses}.
The immediate selection for high stability of the constant 
temperature parameter approach locks the proteins into 
stable structures that cannot easily improve their binding energy,
whereas gradual selection for
stability allows the proteins to first achieve strong binding
and then optimize their stability.  These results indicate that it is easier
to improve stability while maintaining strong binding than it is to improve binding 
while maintaining high stability.  

\subsection*{Evolution from Different Initial Proteins}

The results we have described thus far are from evolutionary trajectories that
began with random protein sequences.  Biological and laboratory 
protein evolution do not start with random sequences, but instead
modify the properties of existing proteins.  To test whether
the trends we observed depend on the initial populations, we
repeated our experiments beginning with proteins that had been
evolved to bind to different ligands by first evolving
random protein populations for $250$ generations to bind to
the three ligands shown in Figure \ref{fig:ligands}.
We then used the most fit protein from each of these runs
as the beginning sequence for $50$ runs of
evolution for binding to the original ligand used
above (Figure \ref{fig:BindingExample}).

All of the trends we found by beginning with random populations were preserved
when we started from these evolved proteins.  We again found that
evolutionary trajectories at lower temperature parameters yielded stronger
binding final proteins, and that the gradient was
more effective at evolving high--temperature fitness than a 
constant high temperature parameter (Figure \ref{fig:diffstarts}).
The different initial populations do lead to different 
final binding energies and fitnesses, with random populations tending
to lead to better values.  However, the trends of lower temperature parameter
leading to stronger binding and a gradient approach leading
to higher fitness hold for all four initial populations.

\section*{Discussion}

We have shown that strong selection for stability
inhibits the evolution of ligand binding function by the proteins in our model.
The ability of a few proteins to evolve strong binding at high
temperature parameters shows that there are sequences
that exhibit both good stability and strong binding. 
However, at high temperature parameters, the evolving proteins are more likely
to become trapped in regions of sequence space that correspond
to highly stable but weakly binding proteins.
Presumably this trapping is due to that fact that stronger selection
for stability reduces the network of sequences that are essentially
neutral since there are fewer highly stable sequences compatible
with a given fold~\citep{Shakhnovich1998,Levitt2002}.   Therefore, at high temperature parameters,
mutations that increase binding are
more likely to lead to an unacceptably large drop in stability.  The net
effect is a roughening of the fitness landscape that makes it more difficult to
escape local optima.

We present a strategy to overcome this problem of the evolutionary
trajectories becoming trapped at high stability  
but weak binding proteins.  Performing the initial rounds of evolution at low
a temperature parameter decreases the selection for stability, and so allows the proteins
to more easily find strong binding regions of sequence space.  The temperature parameter
can then be increased, which leads to the selection of more stable sequences.
Our results indicate that this approach is more effective for evolving highly
stable and strong binding proteins than constant selection for both
high stability and strong binding.  This strategy takes advantage
of the fact that it is easier to maintain strong ligand binding while improving stability
than to maintain high stability while improving binding. 

Our results fit into the framework of current theories about the distributions of
proteins in sequence space that has emerged from other lattice protein studies.
These studies have shown that protein structures are coded for by structurally neutral networks spanning
many diverse sequences, and that these networks are structured as superfunnels,
with the most stable sequences also posessing the most connections in the
networks~\citep{Bornberg-Bauer,Broglia1999,Bastolla,Bastolla1999}.  
Our work suggests that a protein evolves
function most effectively when it can freely explore in its structurally neutral network, rather
then when it is trapped in a small number of highly stable sequences.  Our initial relaxation of
the stability requirement facilitates exploration of the structurally neutral network,
and once highly functional sequences are found, they can be optimized for stability.
Although we do not consider recombination in our current study, other work~\citep{Cui,Xia}
has shown that while structurally neutral networks can easily be explored locally by
point mutations, moves between networks or to distant regions of the same network are
factilited by crossover--induced sequence space jumps.  Therefore, we suggest
that addition of recombination to our evolutionary protocol may further assist in
the evolution of function.

The evolution of our model proteins also has strong parallels with real
protein evolution.  As with real proteins, our model proteins evolve
primarily by structurally conservative mutations that tinker with the 
contacts in a preserved structural 
scaffold, rather than by mutations that cause wholesale structural changes. 
The interplay between the evolution of
stability and function in our model 
is also reminiscent of real protein evolution; 
for example, in the evolution of new function in TEM-1 $\beta$-lactamase,
gains in function were correlated with drops in stability, followed
by gradual regaining of the lost stability~\citep{Wang}. 

Our model points to general trends that are important
in both natural and experimental protein evolution, where
different structural and functional properties are under different
selection pressures.
Protein evolution involves
concurrent selection for stability and function, and
productive mutations must improve one of these properties without 
excessively damaging the other.  Since most mutations to evolved proteins
will be deleterious to at least one of these properties, strong selection
for both stability and function will limit the number of productive mutations,
and so lead to trapping at local fitness optima.  Protein evolution
therefore occurs most efficiently when the temporary drops
in stability associated with gains in function are
buffered by mild selection for stability.

\small 
\section*{}
JDB is supported by a HHMI predoctoral fellowship.  COW and CA were supported
by the NSF under contract number DEB-9981397.  Part of this work was carried out
at the Jet Propulsion Laboratory, California Institute of Technology, under
a contract with the National Aeronautics and Space Administration.  We thank
Hue Sun Chan, George Somero, Jeff Endelman, and Chris 
Voigt for their helpful comments on
the manuscript.

\pagebreak

\pagebreak

\begin{table}
\centerline{\begin{tabular}[t]{|c|c|c|c|}
\hline
Temperature & $\langle \rm{BE} \rangle$ & $\langle \left. \Delta G_{f} \right|_{T = 1.0} \rangle$ & $\langle \left. \mathcal{F} \right|_{T = 1.0} \rangle$ \\
\hline
0.8 &  -18.47 $\pm$ 0.32 & -1.60 $\pm$ 0.06 & 14.75 $\pm$ 0.29 \\
\hline
0.9 &  -16.99 $\pm$ 0.37 & -1.95 $\pm$ 0.07 & 14.57 $\pm$ 0.32 \\
\hline
1.0 &  -15.78 $\pm$ 0.37 & -2.29 $\pm$ 0.07 & 14.18 $\pm$ 0.32 \\
\hline
1.1 &  -15.08 $\pm$ 0.42 & -2.65 $\pm$ 0.09 & 13.99 $\pm$ 0.37 \\
\hline
1.2 &  -13.78 $\pm$ 0.38 & -2.82 $\pm$ 0.09 & 12.97 $\pm$ 0.32 \\
\hline
\end{tabular}}
\caption{\textbf{Average Binding Energies, Stabilities, and Fitnesses} after $500$
generations.  The stabilities and fitnesses are computed at a reference 
temperature parameter
of $1.0$ to allow comparison.  Means are shown plus/minus their
standard errors.  Values are averages
are of the most fit member of all $50$  populations.} 
\label{tab:stats}
\end{table}

\pagebreak

\begin{figure}
\normalsize
\caption{\label{fig:BindingExample} Lowest energy conformation of a protein 
(at left)
bound to the rigid ligand used in the simulations (at right, shown in bold) in
the lowest energy binding position.
The stability of the protein at a temperature parameter of $T=1.0$ is $\Delta G_{f} = -1.04$,
and the binding energy of the protein to the ligand is $\rm{BE} = -17.90$.}
\end{figure}

\begin{figure}
\caption{Two typical replicates performed at a temperature parameter of $1.0$.  The plots at the top
show the evolution of fitness,
of stability (solid lines), and binding energy (dotted lines) of the most fit 
member of the population.  
Structures at bottom are of the most fit sequence at $10$, $170$, $330$,
and $500$ generations.} 
\label{fig:typical}
\end{figure}

\begin{figure}
\caption{Ligand binding function evolves more efficiently at lower temperature parameters.
The histogram shows the distribution of the best binding energies after
$500$ generations of evolution for all $50$ replicates at each temperature parameter.  Binding
energies are of the most fit member of the population.} 
\label{fig:temphist}
\end{figure}

\begin{figure}
\caption{High temperature parameter fitness evolves more efficiently with a 
gradient than with constant selection at a high temperature parameter.
The histogram shows the distribution of fitnesses
after $500$ generations for evolution
at a constant temperature parameter of $1.2$ and evolution with the gradient from
$0.8$ to $1.2$.  Fitnesses are of the most fit member of the 
population.}
\label{fig:grad_fit}
\end{figure} 

\begin{figure}
\caption{Evolution occurs more efficiently when proteins can first evolve
strong ligand binding function at low temperature parameters.  
Solid lines show the stability-function trajectories 
with a gradient
from $0.8$ to $1.2$, and dotted lines show trajectories
with a constant high temperature parameter of $1.2$.  
At left the final fitnesses are $14.01$ for the gradient
and $10.52$ for the constant temperature parameter; at right,
the values are $13.92$ and $10.60$ respectively. 
Plots are for the most fit member of the population sampled every $10$ 
generations.  Stabilities are computed at a reference temperature parameter of
$T = 1.0$ to allow comparison.}
\label{fig:courses}
\end{figure}

\begin{figure}
\caption{Proteins were evolved for $250$ generations to bind to these 
ligands.  We then
modified the function of these evolved proteins by evolving them to bind
to the original ligand shown in Figure \ref{fig:BindingExample}.}
\label{fig:ligands}
\end{figure}

\begin{figure}
\caption{Lower temperature parameter evolution leads to stronger binding,
and a gradient approach leads to better high--temperature parameter fitness
regardless of the initial population.  At left are the
binding energies for
evolution at the indicated temperature parameter with the indicated
initial populations.
At right are the  
fitnesses at a temperature parameter of $1.2$ with evolution at a constant temperature parameter
or with a gradient.  Values are the mean and standard error for
the most fit member of the population for
all $50$ runs after $500$ generations.} 
\label{fig:diffstarts}
\end{figure}

\renewcommand{\baselinestretch}{1}

\DeclareFixedFont{\extrabold}{\encodingdefault}{cmss}{bx}{\shapedefault}{13pt}

\pagebreak

\begin{figure}[h]
\normalsize
\centerline{\fbox{
\begin{tabular}{cc@{}c@{}c@{}c@{}c@{}c@{}c@{}c@{}c@{}c@{}c@{}c@{}cc}
$\cdot$ && $\cdot$ && $\cdot$ && $\cdot$ && $\cdot$ && $\cdot$ && $\cdot$ && $\cdot$ \\
&& && && && && && && \\
$\cdot$ && $\cdot$ && $\cdot$ && $\cdot$ && \extrabold{A} &\textbf{---}& \extrabold{K} &\textbf{---}& \extrabold{G} && $\cdot$ \\
&& && && && && && $\mathbf{\mid}$ && \\
$\cdot$ && \textcolor{black}{K} &\textcolor{black}{---}& \textcolor{black}{H} &\textcolor{black}{---}& \textcolor{black}{C} &\textcolor{black}{---}& \textcolor{black}{I} &\textcolor{black}{---}& \textcolor{black}{E} && \extrabold{K} && $\cdot$ \\
&& \textcolor{black}{$\mid$} && && && && \textcolor{black}{$\mid$} && $\mathbf{\mid}$ && \\
$\cdot$ && \textcolor{black}{K} &\textcolor{black}{---}& \textcolor{black}{F} && \textcolor{black}{F} & \textcolor{black}{---} & \textcolor{black}{M} &\textcolor{black}{---}& \textcolor{black}{R} && \extrabold{E} && $\cdot$ \\
&& && \textcolor{black}{$\mid$} && \textcolor{black}{$\mid$} &&  && && $\mathbf{\mid}$ && \\
$\cdot$ && \textcolor{black}{E} & \textcolor{black}{---} & \textcolor{black}{F} && \textcolor{black}{M} && \textcolor{black}{W} && \extrabold{D} &\textbf{---}& \extrabold{G} && $\cdot$ \\
&& \textcolor{black}{$\mid$} && && \textcolor{black}{$\mid$} && \textcolor{black}{$\mid$} && $\mathbf{\mid}$ && && \\
$\cdot$ && \textcolor{black}{D} &\textcolor{black}{---}& \textcolor{black}{C} && \textcolor{black}{C} &\textcolor{black}{---}& \textcolor{black}{Y} && \extrabold{H} && $\cdot$ && $\cdot$ \\
&& && && && && && && \\
$\cdot$ && $\cdot$ && $\cdot$ && $\cdot$ && $\cdot$ && $\cdot$ && $\cdot$ && $\cdot$ \\
\end{tabular}}}

\centerline{\Large Figure 1}

\end{figure}

\pagebreak

\begin{figure}[h]
\begin{psfrags}
\psfrag{0}[t][t]{\small 0}
\psfrag{150}[t][t]{\small 150}
\psfrag{300}[t][t]{\small 300}
\psfrag{450}[t][t]{\small 450}
\psfrag{-15}[t][t]{\small -15}
\psfrag{-17}[t][t]{\small -17}
\psfrag{-19}[t][t]{\small -19}
\psfrag{-21}[t][t]{\small -21}
\psfrag{-1}[t][t]{\small -1}
\psfrag{-2}[t][t]{\small -2}
\psfrag{10}[t][t]{\small 10}
\psfrag{14}[t][t]{\small 14}
\psfrag{18}[t][t]{\small 18}
\psfrag{Fitness}[t][t]{Fitness}
\psfrag{Stability}[t][t]{Stability}
\psfrag{Binding Energy}[t][t]{Binding Energy}
\includegraphics[width=1.63in]{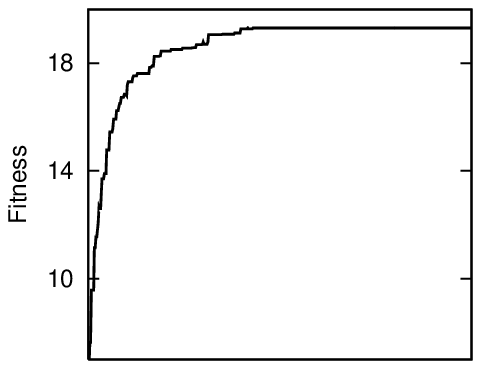}
\includegraphics[width=1.63in]{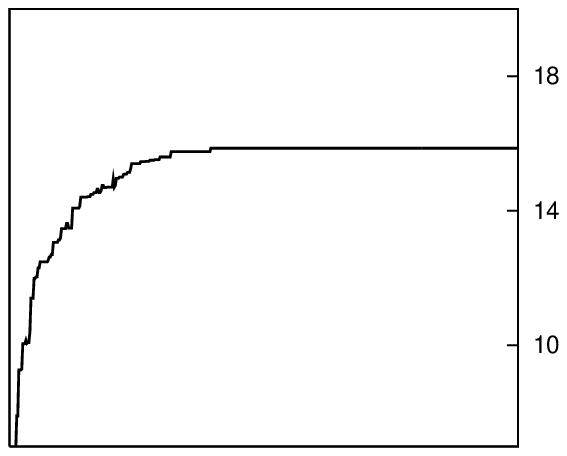}

\includegraphics[width=1.63in]{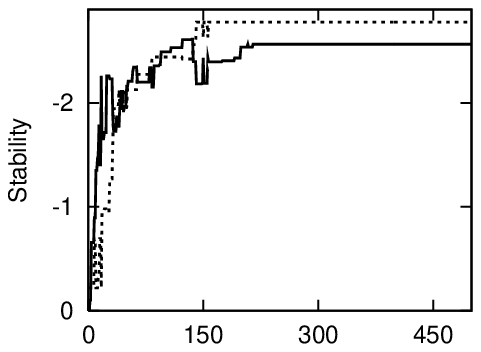}
\includegraphics[width=1.63in]{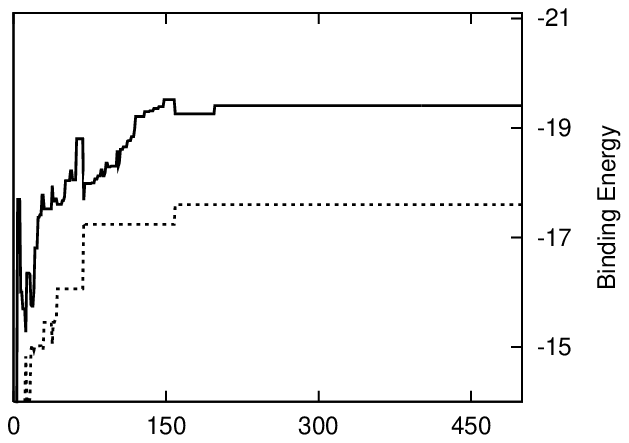}

\centerline{Generation}

\begin{minipage}{1.6in}
\scriptsize
\centerline{
\begin{tabular}{@{}c@{}c@{}c@{}c@{}c@{}c@{}c@{}c@{}c@{}}
$\cdot$          &              & D               & \textbf{---} & A               & \textbf{---}  & Q               &              & $\cdot$         \\
                 &              & $\mathbf{\mid}$ &              &                 &               & $\mathbf{\mid}$ &              &                 \\
$\cdot$          &              & C               &              & I               &               & A               &              & $\cdot$         \\
                 &              &                 &              & $\mathbf{\mid}$ &               & $\mathbf{\mid}$ &              &                 \\
D                & \textbf{---} & F               & \textbf{---} & F               &               & F               & \textbf{---} & R               \\
$\mathbf{\mid}$  &              &                 &              &                 &               &                 &              & $\mathbf{\mid}$ \\
T                & \textbf{---} & V               & \textbf{---} & I               &               & F               & \textbf{---} & N               \\
                 &              &                 &              & $\mathbf{\mid}$ &               & $\mathbf{\mid}$ &              &                 \\
$\cdot$          &              & $\cdot$         &              & S               & \textbf{---}  & D               &              & $\cdot$         \\
\end{tabular}}
\smallskip

\centerline{\Large{$\Downarrow$}}
\smallskip

\centerline{
\begin{tabular}{@{}c@{}c@{}c@{}c@{}c@{}c@{}c@{}c@{}c@{}}
$\cdot$          &              & D               & \textbf{---} & V               & \textbf{---}  & K               &              & $\cdot$         \\
                 &              & $\mathbf{\mid}$ &              &                 &               & $\mathbf{\mid}$ &              &                 \\
$\cdot$          &              & F               &              & I               &               & L               &              & $\cdot$         \\
                 &              &                 &              & $\mathbf{\mid}$ &               & $\mathbf{\mid}$ &              &                 \\
H                & \textbf{---} & F               & \textbf{---} & F               &               & F               & \textbf{---} & K               \\
$\mathbf{\mid}$  &              &                 &              &                 &               &                 &              & $\mathbf{\mid}$ \\
E                & \textbf{---} & F               & \textbf{---} & F               &               & F               & \textbf{---} & K               \\
                 &              &                 &              & $\mathbf{\mid}$ &               & $\mathbf{\mid}$ &              &                 \\
$\cdot$          &              & $\cdot$         &              & C               & \textbf{---}  & K               &              & $\cdot$         \\
\end{tabular}}
\smallskip

\centerline{\Large{$\Downarrow$}} 
\smallskip

\centerline{
\begin{tabular}{@{}c@{}c@{}c@{}c@{}c@{}c@{}c@{}c@{}c@{}}
$\cdot$          &              & D               & \textbf{---} & L               & \textbf{---}  & K               &              & $\cdot$         \\
                 &              & $\mathbf{\mid}$ &              &                 &               & $\mathbf{\mid}$ &              &                 \\
$\cdot$          &              & F               &              & I               &               & L               &              & $\cdot$         \\
                 &              &                 &              & $\mathbf{\mid}$ &               & $\mathbf{\mid}$ &              &                 \\
H                & \textbf{---} & F               & \textbf{---} & F               &               & M               & \textbf{---} & K               \\
$\mathbf{\mid}$  &              &                 &              &                 &               &                 &              & $\mathbf{\mid}$ \\
E                & \textbf{---} & F               & \textbf{---} & M               &               & W               & \textbf{---} & K               \\
                 &              &                 &              & $\mathbf{\mid}$ &               & $\mathbf{\mid}$ &              &                 \\
$\cdot$          &              & $\cdot$         &              & C               & \textbf{---}  & K               &              & $\cdot$         \\
\end{tabular}}
\smallskip

\centerline{\Large{$\Downarrow$}}
\smallskip

\centerline{
\begin{tabular}{@{}c@{}c@{}c@{}c@{}c@{}c@{}c@{}c@{}c@{}}
$\cdot$          &              & D               & \textbf{---} & L               & \textbf{---}  & K               &              & $\cdot$         \\
                 &              & $\mathbf{\mid}$ &              &                 &               & $\mathbf{\mid}$ &              &                 \\
$\cdot$          &              & F               &              & I               &               & L               &              & $\cdot$         \\
                 &              &                 &              & $\mathbf{\mid}$ &               & $\mathbf{\mid}$ &              &                 \\
H                & \textbf{---} & F               & \textbf{---} & F               &               & M               & \textbf{---} & K               \\
$\mathbf{\mid}$  &              &                 &              &                 &               &                 &              & $\mathbf{\mid}$ \\
E                & \textbf{---} & F               & \textbf{---} & M               &               & W               & \textbf{---} & K               \\
                 &              &                 &              & $\mathbf{\mid}$ &               & $\mathbf{\mid}$ &              &                 \\
$\cdot$          &              & $\cdot$         &              & C               & \textbf{---}  & K               &              & $\cdot$         \\
\end{tabular}}
\end{minipage}
\begin{minipage}{0.05in}
\centerline{10}
\vspace{2.8cm}

\centerline{170}
\vspace{2.8cm}

\centerline{330}
\vspace{2.8cm}

\centerline{500}

\end{minipage}
\begin{minipage}{1.6in}
\scriptsize
\centerline{
\begin{tabular}{@{}c@{}c@{}c@{}c@{}c@{}c@{}c@{}c@{}c@{}}
$\cdot$          &              & $\cdot$         &              & S               & \textbf{---}  & D               & &  \\
                 &              &                 &              & $\mathbf{\mid}$ &               & $\mathbf{\mid}$ & &  \\
K                & \textbf{---} & F               & \textbf{---} & V               &               & W               & &  \\
$\mathbf{\mid}$  &              &                 &              &                 &               & $\mathbf{\mid}$ & &  \\
Y                & \textbf{---} & L               &              & L               & \textbf{---}  & T               & &  \\
                 &              & $\mathbf{\mid}$ &              & $\mathbf{\mid}$ &               &                 & &  \\
Y                & \textbf{---} & F               &              & I               & \textbf{---}  & W               & &  \\
$\mathbf{\mid}$  &              &                 &              &                 &               & $\mathbf{\mid}$ & &  \\
P                &              & Y               & \textbf{---} & L               & \textbf{---}  & K               & &  \\
\end{tabular}}
\smallskip

\centerline{\Large{$\Downarrow$}}
\smallskip

\centerline{
\begin{tabular}{@{}c@{}c@{}c@{}c@{}c@{}c@{}c@{}c@{}c@{}}
$\cdot$          &              & $\cdot$         &              & H               & \textbf{---}  & E               & &  \\
                 &              &                 &              & $\mathbf{\mid}$ &               & $\mathbf{\mid}$ & &  \\
E                & \textbf{---} & F               & \textbf{---} & F               &               & I               & &  \\
$\mathbf{\mid}$  &              &                 &              &                 &               & $\mathbf{\mid}$ & &  \\
W                & \textbf{---} & M               &              & F               & \textbf{---}  & P               & &  \\
                 &              & $\mathbf{\mid}$ &              & $\mathbf{\mid}$ &               &                 & &  \\
H                & \textbf{---} & F               &              & W               & \textbf{---}  & M               & &  \\
$\mathbf{\mid}$  &              &                 &              &                 &               & $\mathbf{\mid}$ & &  \\
C                &              & C               & \textbf{---} & M               & \textbf{---}  & K               & &  \\
\end{tabular}}
\smallskip

\centerline{\Large{$\Downarrow$}}
\smallskip

\centerline{
\begin{tabular}{@{}c@{}c@{}c@{}c@{}c@{}c@{}c@{}c@{}c@{}}
$\cdot$          &              & $\cdot$         &              & H               & \textbf{---}  & E               & &  \\
                 &              &                 &              & $\mathbf{\mid}$ &               & $\mathbf{\mid}$ & &  \\
E                & \textbf{---} & F               & \textbf{---} & F               &               & I               & &  \\
$\mathbf{\mid}$  &              &                 &              &                 &               & $\mathbf{\mid}$ & &  \\
W                & \textbf{---} & M               &              & F               & \textbf{---}  & P               & &  \\
                 &              & $\mathbf{\mid}$ &              & $\mathbf{\mid}$ &               &                 & &  \\
H                & \textbf{---} & F               &              & F               & \textbf{---}  & M               & &  \\
$\mathbf{\mid}$  &              &                 &              &                 &               & $\mathbf{\mid}$ & &  \\
C                &              & C               & \textbf{---} & M               & \textbf{---}  & K               & &  \\
\end{tabular}}
\smallskip

\centerline{\Large{$\Downarrow$}}
\smallskip

\centerline{
\begin{tabular}{@{}c@{}c@{}c@{}c@{}c@{}c@{}c@{}c@{}c@{}}
$\cdot$          &              & $\cdot$         &              & H               & \textbf{---}  & E               & &  \\
                 &              &                 &              & $\mathbf{\mid}$ &               & $\mathbf{\mid}$ & &  \\
E                & \textbf{---} & F               & \textbf{---} & F               &               & I               & &  \\
$\mathbf{\mid}$  &              &                 &              &                 &               & $\mathbf{\mid}$ & &  \\
W                & \textbf{---} & M               &              & F               & \textbf{---}  & P               & &  \\
                 &              & $\mathbf{\mid}$ &              & $\mathbf{\mid}$ &               &                 & &  \\
H                & \textbf{---} & F               &              & F               & \textbf{---}  & M               & &  \\
$\mathbf{\mid}$  &              &                 &              &                 &               & $\mathbf{\mid}$ & &  \\
C                &              & C               & \textbf{---} & M               & \textbf{---}  & K               & &  \\
\end{tabular}}
\end{minipage}
\label{fig:typical}
\end{psfrags}

\centerline{\Large Figure 2}

\end{figure}

\newpage

\begin{figure}[h]
\begin{psfrags}
\psfrag{Runs}[t][t]{\small Number of replicates}
\psfrag{0}[t][t]{\footnotesize 0}
\psfrag{6}[t][t]{\footnotesize 6}
\psfrag{12}[t][t]{\footnotesize 12}
\psfrag{18}[t][t]{\footnotesize 18}
\psfrag{24}[t][t]{\footnotesize 24}
\psfrag{30}[t][t]{\footnotesize 30}
\psfrag{Binding Energy}[t][t]{\small Binding energy}
\psfrag{> -12}[t][t]{\footnotesize $>$ -12}
\psfrag{-12 to -14}[t][t]{\footnotesize -12 to -14}
\psfrag{-14 to -18}[t][t]{\footnotesize -14 to -18}
\psfrag{-18 to -20}[t][t]{\footnotesize -18 to -20}
\psfrag{< -19}[t][t]{\footnotesize $<$ -20}
\psfrag{T = 0.8}[t][t]{\footnotesize $T = 0.8$}
\psfrag{T = 0.9}[t][t]{\footnotesize $T = 0.9$}
\psfrag{T = 1.0}[t][t]{\footnotesize $T = 1.0$}
\psfrag{T = 1.1}[t][t]{\footnotesize $T = 1.1$}
\psfrag{T = 1.2}[t][t]{\footnotesize $T = 1.2$}
\centerline{\includegraphics[width=3.25in]{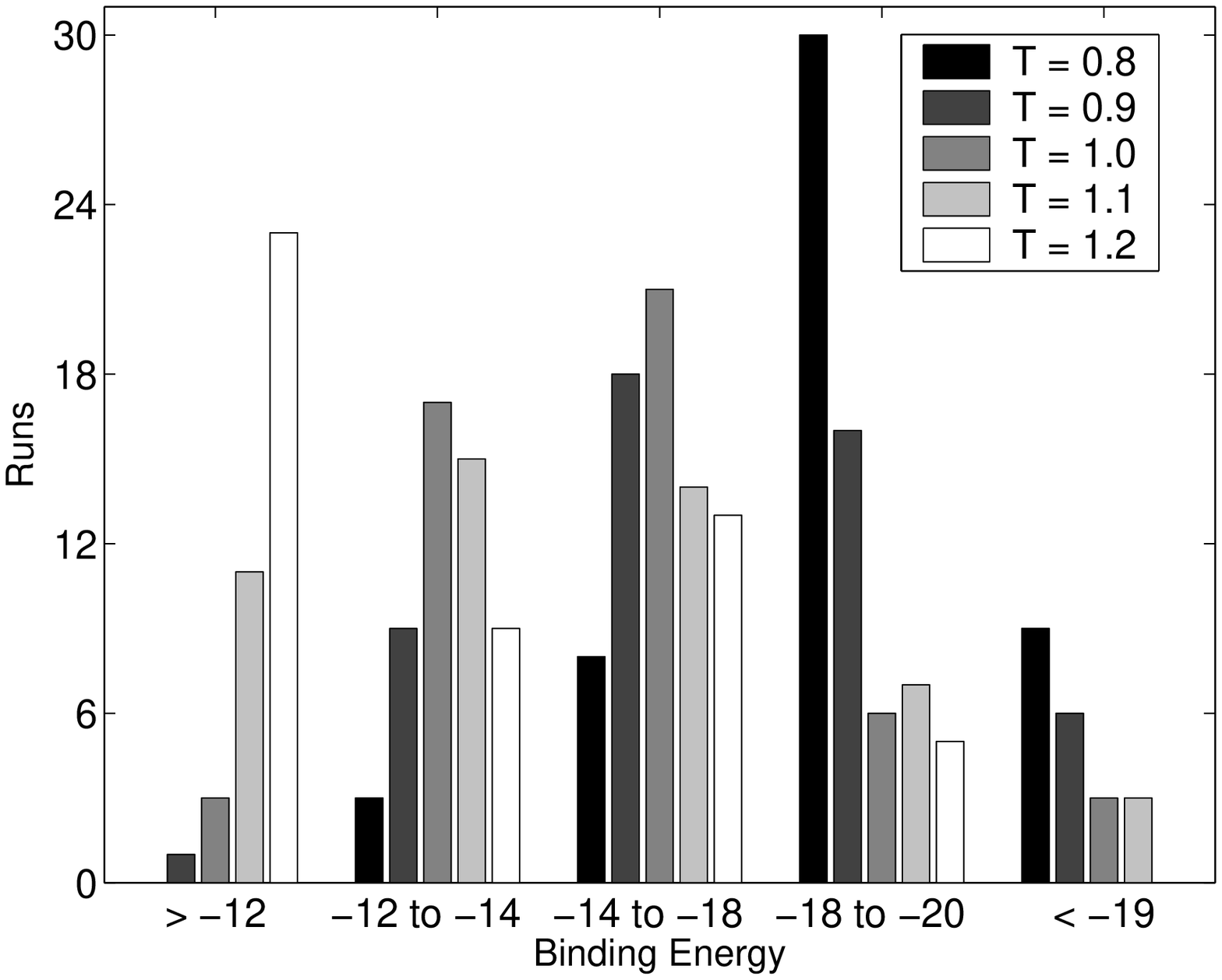}}
\end{psfrags}
\label{fig:temphist}

\centerline{\Large Figure 3}

\end{figure}

\newpage

\begin{figure}[h]
\begin{psfrags}
\psfrag{Fitness}[t][t]{Fitness} 
\psfrag{Runs}[t][t]{Number of replicates}
\psfrag{0}[t][t]{\footnotesize 0}
\psfrag{5}[t][t]{\footnotesize 5}
\psfrag{10}[t][t]{\footnotesize 10}
\psfrag{15}[t][t]{\footnotesize 15}
\psfrag{20}[t][t]{\footnotesize 20}
\psfrag{< 9}[t][t]{\footnotesize $<$ 9}
\psfrag{9 to 11}[t][t]{\footnotesize 9 to 11}
\psfrag{11 to 13}[t][t]{\footnotesize 11 to 13}
\psfrag{13 to 15}[t][t]{\footnotesize 13 to 15}
\psfrag{> 15}[t][t]{\footnotesize $>$ 15}
\psfrag{constant}[t][t]{\footnotesize constant}
\psfrag{gradient}[t][t]{\footnotesize gradient}
\centerline{\includegraphics[width=3.25in]{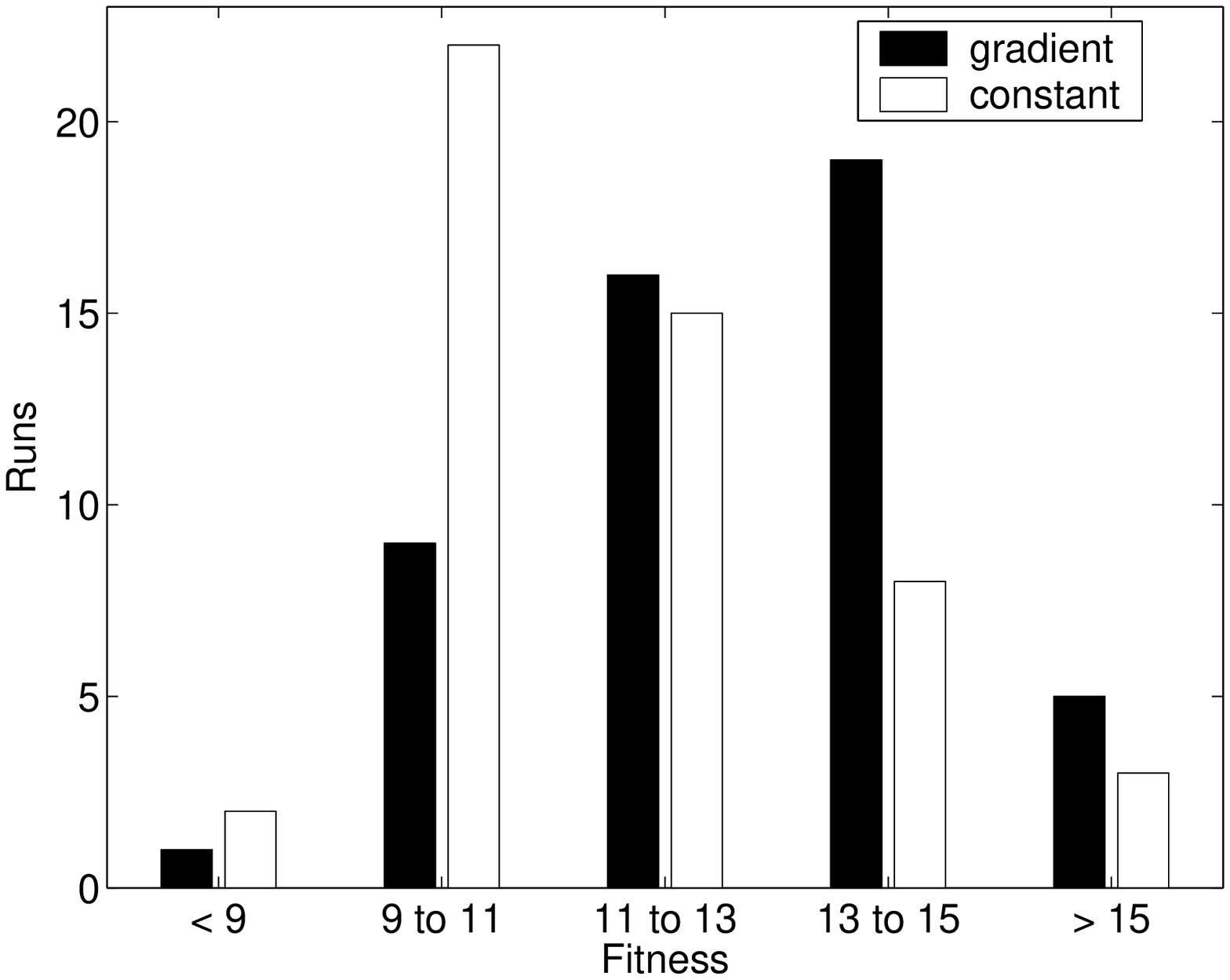}}
\end{psfrags}
\label{fig:grad_fit}

\centerline{\Large Figure 4}

\end{figure}

\newpage

\begin{figure}[h]
\begin{psfrags}
\psfrag{Binding Energy}[t][t]{\small Binding Energy}
\psfrag{0}[t][t]{\small 0}
\psfrag{-1}[t][t]{\small -1}
\psfrag{-2}[t][t]{\small -2}
\psfrag{-3}[t][t]{\small -3}
\psfrag{-15}[t][t]{\small -15}
\psfrag{-20}[t][t]{\small -20}
\psfrag{-10}[t][t]{\small -10}
\psfrag{-14}[t][t]{\small -14}
\psfrag{-18}[t][t]{\small -18}
\begin{minipage}{1.7in}
\includegraphics[width=1.7in]{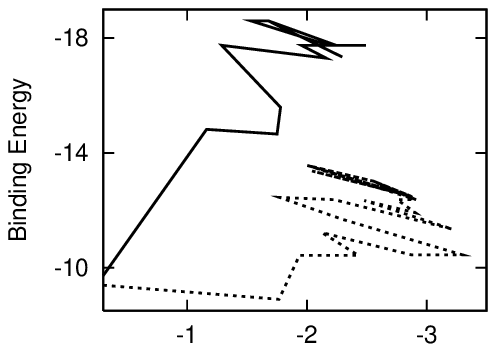}
\end{minipage}
\begin{minipage}{1.7in}
\includegraphics[width=1.7in]{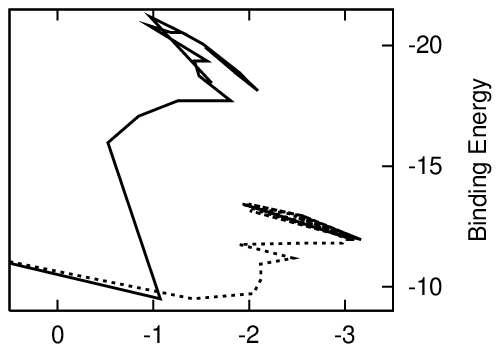}
\end{minipage}
\end{psfrags}
\centerline{\small Stability (at $T = 1.0$)}
\label{fig:courses}

\centerline{\Large Figure 5}

\end{figure}

\newpage

\begin{figure}[h]
\begin{minipage}[b]{1.0in}
\footnotesize
\centerline{
\begin{tabular}{@{}c@{}c@{}c@{}}
A                & \textbf{---} & W               \\
                 &              & $\mathbf{\mid}$ \\
K                & \textbf{---} & Y               \\
$\mathbf{\mid}$  &              &                 \\
S                &              & $\cdot$         \\
\end{tabular}}
\normalsize
\centerline{Ligand 1}
\end{minipage}
\begin{minipage}[b]{1.4in}
\footnotesize
\centerline{
\begin{tabular}{@{}c@{}c@{}c@{}c@{}c@{}c@{}c@{}c@{}c@{}}
$\cdot$         &               & $\cdot$        &               & $\cdot$         &               & $\cdot$    &               & H               \\
                &               &                &               &                 &               &            &               & $\mathbf{\mid}$ \\
$\cdot$         &               & $\cdot$        &               & $\cdot$         &               & $\cdot$    &               & D               \\
                &               &                &               &                 &               &            &               & $\mathbf{\mid}$ \\
V               & \textbf{---}  & H              & \textbf{---}  & P               & \textbf{---}  & A          & \textbf{---}  & G               \\
$\mathbf{\mid}$ &               &                &               &                 &               &            &               &                 \\
L               &               & $\cdot$        &               & $\cdot$         &               & $\cdot$    &               & $\cdot$         \\
$\mathbf{\mid}$ &               &                &               &                 &               &            &               &                 \\
G               &               & $\cdot$        &               & $\cdot$         &               & $\cdot$    &               & $\cdot$         \\
\end{tabular}}
\normalsize
\centerline{Ligand 2}
\end{minipage}
\begin{minipage}[b]{1.4in}
\footnotesize
\centerline{
\begin{tabular}{@{}c@{}c@{}c@{}c@{}c@{}c@{}c@{}c@{}c@{}c@{}c@{}}
V               & \textbf{---}  & S              & \textbf{---}  & E               & \textbf{---}  & W               &               & $\cdot$         &      & $\cdot$   \\
$\mathbf{\mid}$ &               &                &               &                 &               & $\mathbf{\mid}$ &               &                 &      &           \\
L               &               & $\cdot$        &               & $\cdot$         &               & H               & \textbf{---}  & G               &      & $\cdot$   \\
                &               &                &               &                 &               &                 &               & $\mathbf{\mid}$ &      &           \\
$\cdot$         &               & $\cdot$        &               & $\cdot$         &               & $\cdot$         &               & K               & \textbf{---} & E \\
                &               &                &               &                 &               &                 &               &                 & & $\mathbf{\mid}$ \\
$\cdot$         &               & $\cdot$        &               & $\cdot$         &               & $\cdot$         &               & R               & \textbf{---} & K \\
\end{tabular}}
\normalsize
\centerline{Ligand 3}
\end{minipage}
\label{fig:ligands}

\centerline{\Large Figure 6}

\end{figure}

\newpage

\begin{figure}[h]
\begin{psfrags}
\psfrag{-12}[t][t]{\footnotesize{-12}}
\psfrag{-14}[t][t]{\footnotesize{-14}}
\psfrag{-16}[t][t]{\footnotesize{-16}}
\psfrag{-18}[t][t]{\footnotesize{-18}}
\psfrag{0.8}[t][t]{\footnotesize{0.8}}
\psfrag{0.9}[t][t]{\footnotesize{0.9}}
\psfrag{1.0}[t][t]{\footnotesize{1.0}}
\psfrag{1.1}[t][t]{\footnotesize{1.1}}
\psfrag{1.2}[t][t]{\footnotesize{1.2}}
\psfrag{10}[t][t]{\footnotesize{10}}
\psfrag{11}[t][t]{\footnotesize{11}}
\psfrag{12}[t][t]{\footnotesize{12}}
\psfrag{13}[t][t]{\footnotesize{13}}
\psfrag{constant}[t][t]{\small{constant}}
\psfrag{gradient}[t][t]{\small{gradient}}
\psfrag{Ligand 1}[t][t]{\footnotesize{Ligand 1}}
\psfrag{Ligand 2}[t][t]{\footnotesize{Ligand 2}}
\psfrag{Ligand 3}[t][t]{\footnotesize{Ligand 3}}
\psfrag{Random}[t][t]{\footnotesize{Random}}
\psfrag{Temperature}[t][t]{\small{Temperature}}
\psfrag{Binding Energy}[t][t]{\small{Binding Energy}}
\psfrag{Fitness}[t][t]{\small{Fitness}}
\includegraphics[width=2.2in]{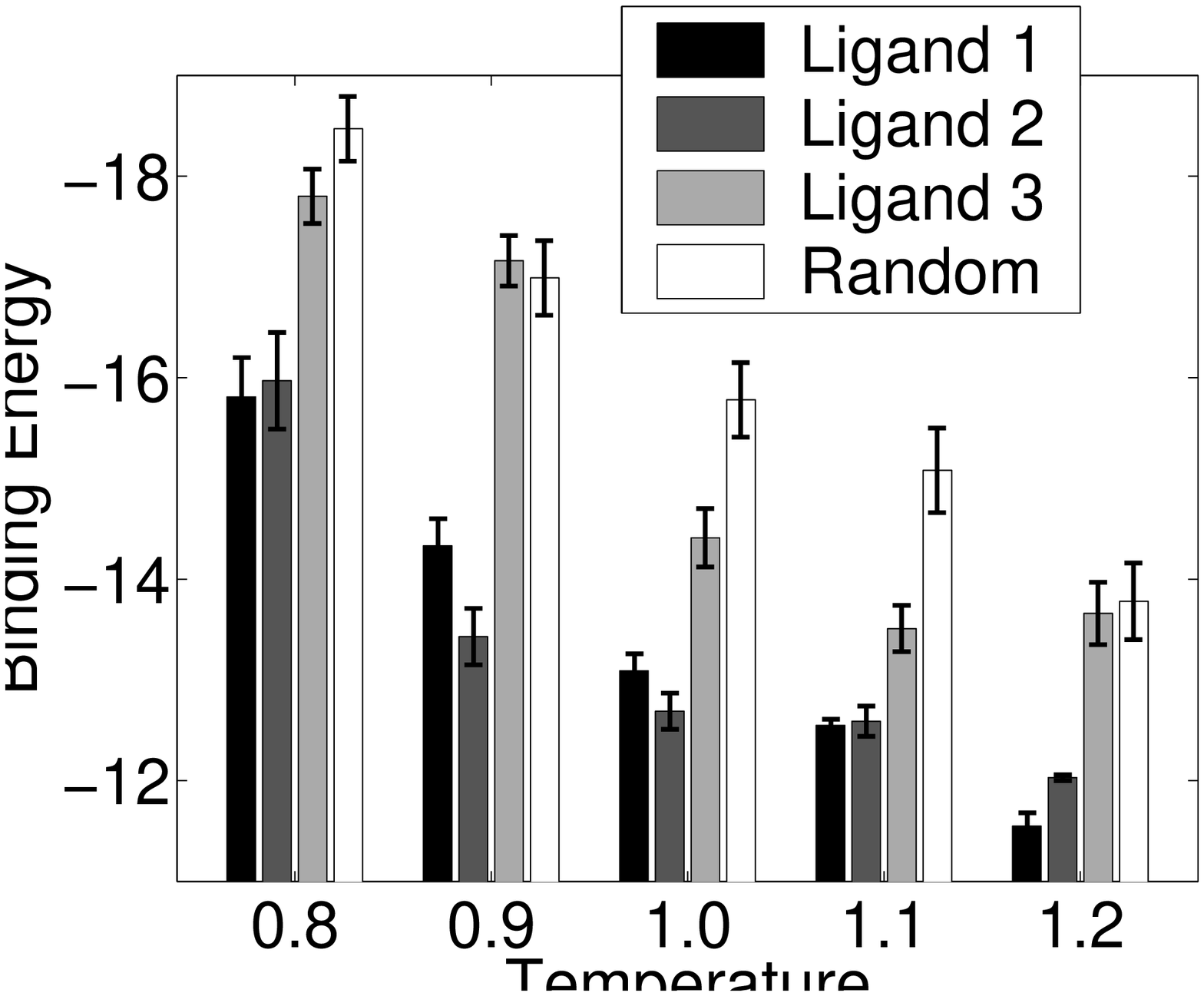}
\includegraphics[width=2.2in]{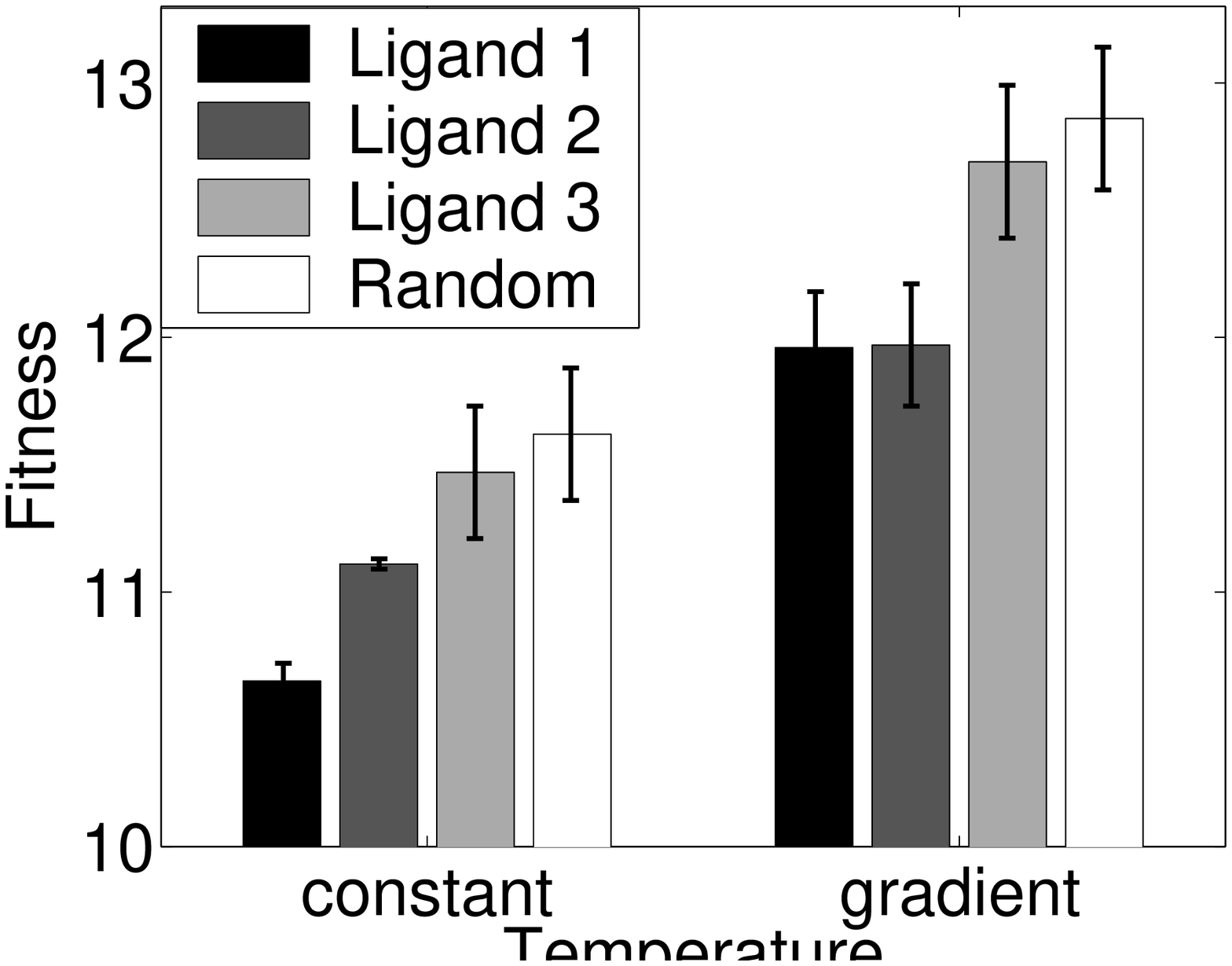}
\end{psfrags}
\label{fig:diffstarts}

\centerline{\Large Figure 7}

\end{figure}

\end{document}